\documentclass[%
 reprint,
 amsmath,amssymb,
 aps,
]{revtex4-2}

\usepackage{graphicx}
\usepackage{dcolumn}
\usepackage{bm}
\usepackage[colorlinks=true,bookmarks=false,citecolor=blue,urlcolor=blue,linkcolor=black]{hyperref} 
\usepackage{array}
\usepackage{tabularx}
\usepackage{verbatim}
\usepackage{cellspace}
\usepackage{tabularray}
\usepackage{textgreek} 
\usepackage{upgreek} 

\begin{document}

\title{
Full-wave simulations of tomographic optical imaging inside scattering media
}

\author{Zeyu Wang}
\author{Yiwen Zhang}
\author{Chia Wei Hsu}
 \email{cwhsu@usc.edu}
\affiliation{%
Ming Hsieh Department of Electrical and Computer Engineering, University of Southern California, Los Angeles, California 90089, USA
}%

\begin{abstract}
Label-free tomographic optical imaging inside scattering media is important for medical diagnosis, biological science, colloidal physics, and device inspection.
An outstanding challenge is that the ground-truth structure is often unknown, so one cannot rigorously assess and compare different imaging schemes.
Here we demonstrate full-wave simulations of tomographic optical imaging deep inside scattering media, which
provide not only the ground truth but also the flexibility to tailor the structure and the convenience of comparing different imaging schemes in the same virtual setup with minimal cost.
We model reflectance confocal microscopy, optical coherence tomography, optical coherence microscopy, interferometric synthetic aperture microscopy, and the recently proposed scattering matrix tomography for imaging nanoparticle targets embedded in a large scattering medium. 
The ground truth enables the identification of artifacts that would typically be mistaken as being correct while setting a rigorous and uniform standard across different methods.
This work shows how full-wave simulations can fill the gaps of experiments for studying imaging inside scattering media.
\end{abstract}

\maketitle

\section{Introduction}

Light scattering from inhomogeneity limits the imaging depth and resolution~\cite{Yoon2020,Bertolotti2022}.
To filter away multiple scattering from the signal, one may adopt spatial gating as in reflectance confocal microscopy (RCM)~\cite{Que2015}, time/coherence gating as in optical coherence tomography (OCT)~\cite{Huang1991}, or a combination of both as in optical coherence microscopy (OCM)~\cite{Izatt1994, Beaurepaire1998}. 
The interferometric synthetic aperture microscopy (ISAM)~\cite{Ralston2007} improves the degradation of resolution away from the focal plane in OCT and OCM.
Matrix-based iterative phase conjugation~\cite{Kang2017} and singular value decomposition~\cite{Badon2016, Badon2020} were developed to correct for aberrations and scattering at individual focal planes.
Optical diffraction tomography (ODT) solves an inverse problem to reconstruct the refractive index profile of the system~\cite{Jin2017, 2019_Lim_LSA, 2019_Chowdhury_Optica, 2022_Zhu_OE, Chen2020}.
Recently, we proposed and experimentally demonstrated the ``scattering matrix tomography'' (SMT) method~\cite{Hsu2023}, which uses digital spatiotemporal gating and digital wavefront corrections to enable high-resolution imaging at deep depths across a large volume. 

To improve these imaging methods, one should first determine when and how they fail, but rigorous assessments are typically limited to prefabricated planar targets where the ground truth is known.
For tomographic reconstruction, the imaging performance can only be inferred from indirect metrics such as visual appearance and image sharpness except with 3D-printed targets~\cite{2022_Krauze_srep} which come with restrictions on the feature size and topology of the structure.
The indirect metrics cannot reveal, for example, when an artifact is produced in the image or when some targets are missed in the image.
Additionally, comparing different methods requires building multiple setups while imaging the identical volume of the sample, which is difficult, time-consuming, and costly since the different methods require different measurements.

Numerical simulations can provide the ground truth, as well as the flexibility to tailor the structure and the convenience of testing different imaging schemes in the same virtual setup.
Label-free scattering-based imaging methods like RCM, OCT/OCM/ISAM, ODT, and SMT are all rigorously governed by Maxwell's equations. 
However, solving Maxwell's equations at the macroscopic length scales in the multiple-scattering regime requires significant computing resources, and the necessity of an angular scan or spatial scan of the incident wave for image formation makes such modeling prohibitive.
Multi-slice or multi-layer models divide the system into single-scattering layers to reduce computing cost but generally do not account for reflection~\cite{2019_Lim_LSA, 2019_Chowdhury_Optica, 2022_Zhu_OE}; the versions that treat reflection~\cite{Chen2020} still ignore multiple reflections and multiple scattering within each layer (which can accumulate across layers).
The extended Huygens--Fresnel formalism~\cite{1997_Schmitt_JOSAA} and Monte Carlo methods~\cite{1999_Yao_PMB} ensemble average over the disorder and neglect the coherence of the multiple-scattering waves.
Prior full-wave computations were limited to a single incident wave with no scanning~\cite{2015_Tseng_BOE, 2018_Kim_BOE} which is insufficient for imaging, simulations of OCT in small systems in the single-scattering~\cite{2010_Reed_SPIE, 2012_Capoglu_book_chapter, 2015_Huang_SPIE, Munro2016, Brenner2019} or few-scattering regime~\cite{Munro2015, 2021_Macdonald_BOE}, and simulations of diffraction tomography in small systems a few wavelengths in size~\cite{2018_Liu_TCI}.
All of these approaches require one simulation per incident wave and are ill-suited for modeling imaging experiments, which typically involve thousands of inputs or more.

Here, we realize full-wave simulations that model SMT, RCM, OCT, OCM, and ISAM in a large-scale 2D system deep in the multiple-scattering regime. 
We develop the source and detection schemes that model these imaging methods in a unified virtual setup and overcome the computational challenge by adopting a new numerical approach ``augmented partial factorization'' (APF)~\cite{Hsu2022}.
APF offers orders-of-magnitude of speedup by handling all incident waves of interest simultaneously and by directly evaluating the quantities measured in an imaging experiment ({\it i.e.}, signal at the detectors in the far field) without solving for the volumetric field profile in the interior of the sample.
Knowing the ground truth enables us to (1) validate the numerical modeling and the image reconstruction algorithms using a weakly scattering system where all imaging methods are expected to work, and (2) consider a strongly scattering system and identify false-positive artifacts in the images and false-negative missing targets that arise from multiple scattering and would have gone unnoticed without knowledge of the ground truth.
With the unified virtual setup, we perform a quantitative comparison between different imaging methods and validate the capabilities of the recently proposed SMT. 
The computations can run on a single desktop or laptop, and we have made our simulation and image reconstruction codes open-source~\cite{GitHub_imaging_simulations}.
Such full-wave simulations can complement experiments and provide a versatile toolbox to explore imaging in scattering media.

\vspace{-3pt}
\section{Full-wave modeling}
\vspace{-2pt}

In this section, we describe how we model the different imaging methods with full-wave simulations.

\vspace{-6pt}
\subsection{Imaging methods}\label{sec:imaging_methods}
\vspace{-4pt}

\begin{figure}
    \includegraphics[width=0.45\textwidth]{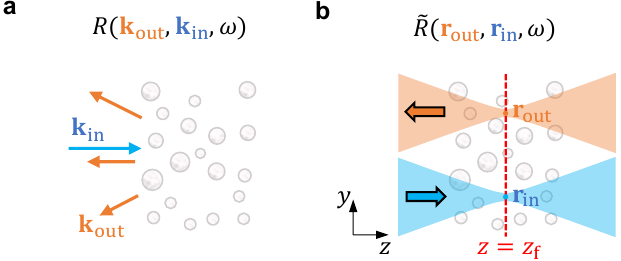}
    \caption{\label{fig:r_schematic}{\bf Quantities computed from the full-wave simulations.}
    {\bf a}, The angular reflection matrix $R({\bf k}_{\rm out}, {\bf k}_{\rm in}, \omega)$.
    {\bf b}, The spatial reflection matrix $\tilde{R}({\bf r}_{\rm out}, {\bf r}_{\rm in}, \omega)$ at focal plane $z=z_{\rm f}$.}
\end{figure}

We start by defining the quantities that need to be computed for each of these imaging methods.
SMT~\cite{Hsu2023} in reflection mode reconstructs a volumetric image of the sample using its hyperspectral reflection matrix $R({\bf k}_{\rm out}, {\bf k}_{\rm in}, \omega)$, which is the field amplitude scattered into direction with momentum $\bf k_{\rm out}$ given an incident plane wave with momentum $\bf k_{\rm in}$ at frequency $\omega$ (Fig.~\ref{fig:r_schematic}{\bf a}). 
The idea is to digitally synthesize an incident pulse
$E_{\rm in}(\bf r, \it t) = \sum_{\bf k_{\rm in}, \omega} e^{i {\bf k}_{\rm in} \cdot ({\bf r} - {\bf r}_{\rm in}) - i \omega (t-t_0)}$
that focuses to position ${\bf r} = \bf r_{\rm in}$ at time $t=t_0$. 
By evaluating the resulting scattered field $E_{\rm out}({\bf r}_{\rm out}, \it t)$, aligning the input spatial gate with the output spatial gate (setting ${\bf r}_{\rm in} = {\bf r}_{\rm out} = {\bf r}$), aligning the temporal gate with the spatial gates (setting $t=t_0$), and digitally scanning ${\bf r}$, the SMT image is formed through a non-uniform Fourier transform~\cite{Hsu2023}:
\begin{equation}
    I_{\rm SMT}({\bf r}) = \left| \sum_{{\bf k}_{\rm out}, {\bf k}_{\rm in}, \omega}
    e^{i({\bf k}_{\rm out}-{\bf k}_{\rm in})\cdot{\bf r}}
    R({\bf k}_{\rm out}, {\bf k}_{\rm in}, \omega) \right|^2.
    \label{eq:psi_smt}
\end{equation}
The spatial and chromatic aberrations created by the refractive index mismatch between interfaces~\cite{1993_Hell} is digitally corrected using the appropriate reference plane and the appropriate propagation phase shift $(\bf k_{\rm out}-\bf k_{\rm in})\cdot{\bf r}$ in each medium.
For simplicity, here we skip the additional wavefront optimizations and corrections introduced in Ref.~\cite{Hsu2023}.
To model SMT, we need to compute the hyperspectral reflection matrix $R({\bf k}_{\rm out}, {\bf k}_{\rm in}, \omega)$ with angles within the numerical aperture (NA) of the objective lens.

RCM operates with a monochromatic light (with $\omega=\omega_{\rm RCM }$) and adopts a spatial focus in detection that coincides with the spatial focus in illumination.
A volumetric image is formed by scanning the focus~\cite{Que2015}:
\begin{equation}
    I_{\rm RCM}({\bf r}) = \left| \tilde{R}({\bf r}_{\rm out}={\bf r}, {\bf r}_{\rm in} = {\bf r}, \omega=\omega_{\rm RCM})
    \right|^2.
    \label{eq:psi_rcm_exp}
\end{equation}
Here, we use $\tilde{R}({\bf r}_{\rm out}, {\bf r}_{\rm in}, \omega)$ to denote the reflection matrix in spatial basis where the output and the input are beams focused to positions ${\bf r}_{\rm out}$ and ${\bf r}_{\rm in}$ respectively in the absence of scattering (Fig.~\ref{fig:r_schematic}{\bf b}). Direct modeling of RCM translates to one simulation for every point ${\bf r}$ of the image. 
To lower the computing cost, here we synthesize the input and output spatial gates using the angular reflection matrix at frequency $\omega_{\rm RCM }$, such that
RCM reduces to SMT at a single frequency,
\begin{equation}
    I_{\rm RCM}({\bf r})= \left| \sum_{\bf k_{\rm out}, \bf k_{\rm in} } e^{i ({\bf k}_{\rm out} -{\bf k}_{\rm in})\cdot{\bf r}} R(\bf k_{\rm out}, \bf k_{\rm in}, \omega = \omega_{\rm RCM}) \right|^2,
    \label{eq:psi_rcm}
\end{equation}
without time gating and without the index mismatch correction.
No additional quantities need to be computed from simulations.

OCT and OCM~\cite{Huang1991,Izatt1994, Beaurepaire1998} adopt a confocal spatial gate at ${\bf r}_{\rm f} = ({\bf r}_{\parallel}, z_{\rm f})$ where the transverse coordinate ${\bf r}_{\parallel}$ is scanned while the axial focal depth is fixed at the focal plane $z = z_{\rm f}$.
Coherence gating (equivalent to time gating) is used to scan the axial coordinate of the image.
We denote
\begin{equation}
R_{\rm c}({\bf r}_{\parallel}, \omega) \equiv \tilde{R}({\bf r}_{\rm out}={\bf r}_{\rm f}, {\bf r}_{\rm in} = {\bf r}_{\rm f}, \omega)
\end{equation}
as the hyperspectral confocal reflection coefficient at the focal plane, with ${\bf r}_{\rm f} = ({\bf r}_{\parallel}, z_{\rm f})$.
Then the temporal gate for depth $z$ requires an additional round-trip time of
$\Delta t = 2(z-z_{\rm f})/v_{\rm g}$ where $v_{\rm g}$ is the group velocity in the medium.
This yields the OCT or OCM image as
\begin{equation}
    I_{\rm OCT/OCM}({\bf r}_{\parallel}, z) = \left| \sum_{\omega}
     e^{-2i\omega\frac{z - z_{\rm f}}{v_{\rm g}}}
    R_{\rm c}({\bf r}_{\parallel}, \omega)
    \right|^2.
    \label{eq:psi_ocm_or_oct}
\end{equation}
To model frequency-domain OCT and OCM with simulations, we need to compute $R_{\rm c}({\bf r}_{\parallel}, \omega)$, which are the diagonal elements of the spatial reflection matrix (Fig.~\ref{fig:r_schematic}{\bf b}).
OCT adopts a weakly focused illumination and detection with a low NA, which provides a relatively large depth of field (equaling twice the Rayleigh range) but a low lateral resolution.
OCM adopts a high NA for higher lateral resolution, resulting in a smaller depth of field.

ISAM uses inverse scattering to extend the depth of field of OCT and OCM~\cite{Ralston2007, Adie2015}.
It uses the same confocal reflection coefficient data $R_{\rm c}({\bf r}_{\parallel}, \omega)$ as OCT/OCM, but instead of performing the temporal gate directly, it performs volumetric image reconstruction in the spatial frequency ${\bf q} = {\bf k}_{\rm out} - {\bf k}_{\rm in} = ({\bf q}_{\parallel}, q_z)$ coordinate.
First, a transverse Fourier transform converts the transverse position ${\bf r}_{\parallel}$ to the transverse spatial frequency ${\bf q}_{\parallel}$,
$\bar{R}({\bf q}_{\parallel}, \omega)=\int R_{\rm c}({\bf r}_{\parallel}, \omega)e^{-i {\bf q}_{\parallel} \cdot {\bf r}_{\parallel}}d{\bf r}_{\parallel}$.
Each ${\bf q}_{\parallel}$ term here arises from summing many different pairs of ${\bf k}_{\rm out}$ and ${\bf k}_{\rm in}$ in the detection and illumination, each with its own $q_z = k_z^{\rm out} - k_z^{\rm in}$.
ISAM assigns the same $q_z$ to all such pairs by considering 
the pair with ${\bf k}_{\rm out} = -{\bf k}_{\rm in}$, where ${\bf q} = 2{\bf k}_{\rm out}$ and $q_z = 2 k_z^{\rm out} = -2 \sqrt{(n_{\rm eff}\omega/c)^2 - |{\bf q}_{\parallel}/2|^2}$, with $n_{\rm eff}$ being the effective index of the medium.
This choice is justified with a geometric optics argument by considering scatterers that are far from the focal plane, which are the ones that suffer the most from defocus of the illumination and detection~\cite{Adie2015}.
Under this so-called ``far-from-focus'' condition,
each element of $\bar{R}({\bf q}_{\parallel}, \omega)$ maps to one element of ${R}_{\rm ISAM}({\bf q}_{\parallel}, q_z) = {R}_{\rm ISAM}({\bf q})$.
An inverse Fourier transform from ${\bf q}$ to ${\bf r}$ forms the ISAM image,
\begin{equation}
    I_{\rm ISAM}({\bf r}) = \left| \sum_{{\bf q}}
    e^{i{\bf q} \cdot {\bf r}} {R}_{\rm ISAM}({\bf q}) \right|^2.
    \label{eq:psi_isam}
\end{equation}
Since ISAM uses the same data $R_{\rm c}({\bf r}_{\parallel}, \omega)$ as OCT/OCM, no additional quantities need to be computed from simulations.

\vspace{-6pt}
\subsection{Governing equation and model system}\label{sec:model_system}
\vspace{-4pt}

\begin{figure*}
    \includegraphics[width=0.995\textwidth]{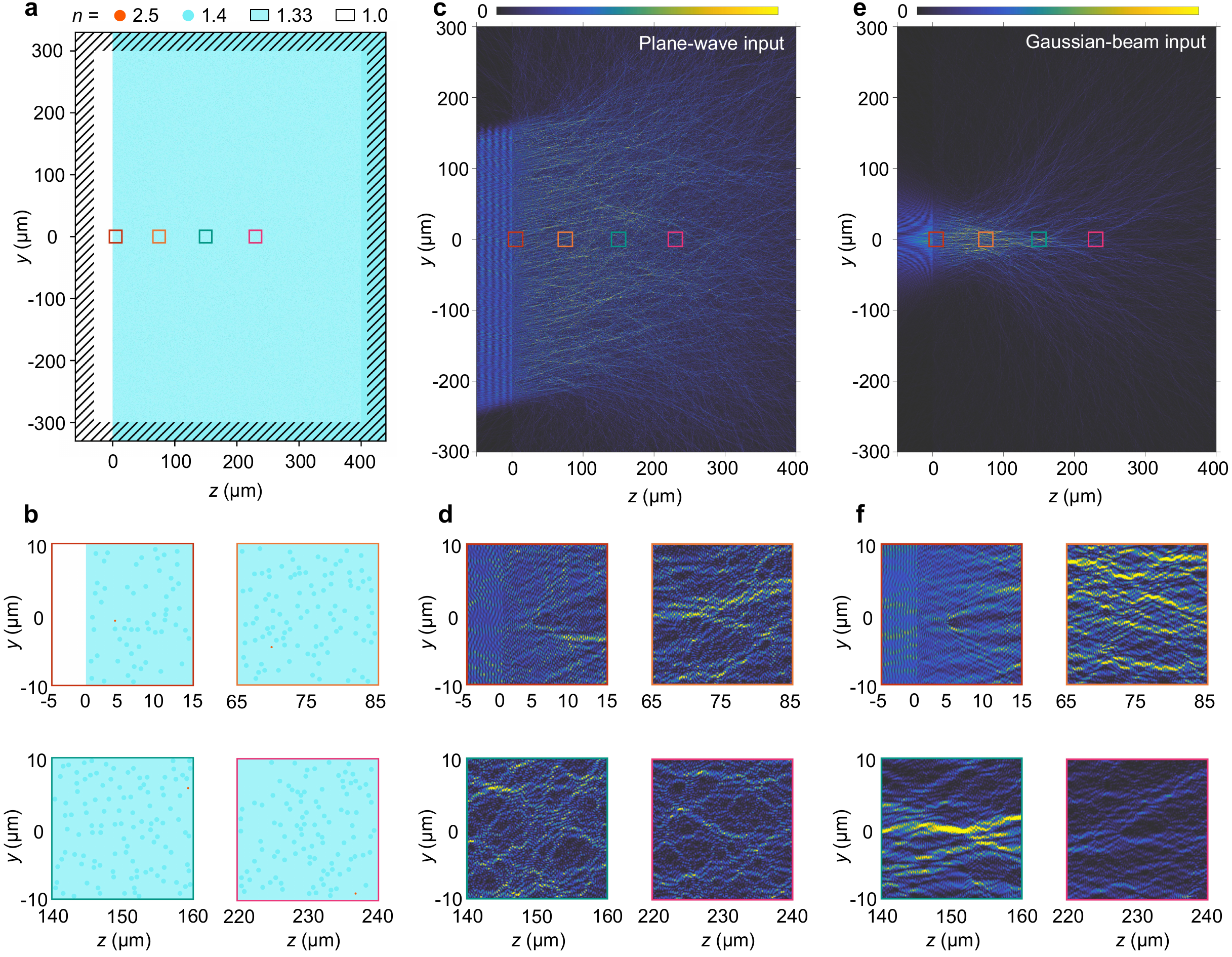}
    \caption{\label{fig:simulation_setup}{\bf Model system and intensity profiles from full-wave simulations.}
    {\bf{a--b}}, Refractive index profile of the system, consisting of a random assembly of 480 high-index (red disks, $n_{\rm high}=2.5$, diameter 300 nm) and 60,000 low-index (blue disks, $n_{\rm low}=1.4$, diameter 700 nm) cylinders embedded in a homogeneous background medium (blue, $n_{\rm bg}=1.33$). Illumination and detection are both placed on the left, in air (white; $n_{\rm air}=1$). The colored square boxes indicate the zoom-in regions shown in {\bf b}. Dashed lines schematically indicate perfectly matched layers. 
    {\bf{c--f}}, Intensity profiles $|E_x({\bf r})|^2$ of the system computed from full-wave simulations under illumination by a truncated plane wave at $15^{\circ}$ incidence in air ({\bf c--d}, for computing the angular reflection matrix) and by 
    a normal incident Gaussian beam with numerical aperture NA = 0.5 focused to $z_{\rm f} = 150$ \textmu m, $y=0$ ({\bf e--f}, for computing the confocal reflection coefficient), at wavelength $\lambda = 823$ nm. }
\end{figure*}

To fully describe the physics of the multiple scattering of light, we model the imaging methods by directly solving Maxwell's equations of the inhomogeneous medium.
We consider the structure and the fields to be independent of the $x$ coordinate, so the equations reduce to a 2D system in ${\bf r} = (y,z)$, and the transverse magnetic (TM) fields $H_y$, $H_z$, $E_x$ are decoupled from the transverse electric (TE) fields $E_y$, $E_z$, $H_x$.
We adopt the TM polarization, governed by
\begin{equation}
\label{eq:wave_eq}
    \left [-\frac{\partial^2}{\partial y^2}-\frac{\partial^2}{\partial z^2}-\frac{\omega^2}{c^2}\varepsilon_{\rm r}({\bf r})
    \right ] E_m^{\rm tot}({\bf r}) = b_m({\bf r}),
\end{equation}
with $b_m({\bf r})$ being the $m$-th input source profile and $E_m^{\rm tot}$ being the $x$ component of the total electric field.
We use our open-source MESTI software~\cite{MESTI_GitHub} to solve Eq.~\eqref{eq:wave_eq} under finite-difference discretization at grid size $\Delta x = 0.1$ \textmu m with subpixel smoothing~\cite{2006_Farjadpour_OL}, at 450 wavelengths $\lambda = 2\pi c/\omega$ from 700 nm to 1000 nm.
We use the finite-difference dispersion relation~\cite{Hsu2022} ${\bf k}(\omega)$ to account for the numerical dispersion from discretization.

With the formalism of Eq.~\eqref{eq:wave_eq}, we can adopt any relative permittivity profile $\varepsilon_{\rm r}({\bf r})$ to describe arbitrarily complex structures.
For simplicity and ease of characterization, here we consider a mixture of two types of scatterers (Fig.~\ref{fig:simulation_setup}a--b):
a dense assembly of low-index-contrast ($n_{\rm low}=1.4$) cylinders with diameter 700 nm (anisotropy factor $g \equiv \langle \cos \theta \rangle \approx 0.89$) that mimics biological tissue but with a stronger scattering strength, and
a sparse collection of high-index ($n_{\rm high}=2.5$) cylinders with diameter 300 nm that act as the targets to be imaged,
in a background medium with refractive index $n_{\rm bg}=1.33$.
The overall scattering mean free path and transport mean free path, accounting for the effect of spatial correlations in the positions of the scatterers, are $l_{\rm sca} \approx 53$ \textmu m and $l_{\rm tr} \approx 350$ \textmu m (Supplementary Sec.~1).
For simplicity, all scatterers here are randomly positioned, and we do not explicitly include fluctuations at longer length scales (which can be modeled by changing the scatterer positions or by changing $\varepsilon_{\rm r}({\bf r})$ directly).
The system size here is $W=600$ \textmu m wide in $y$ and $L=400$ \textmu m thick in $z$, about $850\lambda \times 570\lambda$ at the shortest wavelength.
The effective index of the medium in the long-wavelength limit, evaluated by averaging the relative permittivity in space, is $n_{\rm eff} \approx 1.34$.
We consider far-field illumination and detection both from the left side of the medium, in the air ($n_{\rm air}=1$).
Perfectly matched layers (PMLs)~\cite{2005_Gedney_book_chapter} are placed on all sides of the simulation domain to describe the open boundaries.

\vspace{-6pt}
\subsection{Illumination and detection schemes} \label{sec:input_and_output}
\vspace{-4pt}

As described in Sec.~\ref{sec:imaging_methods}, we need the angular reflection matrix $R({\bf k}_{\rm out}, {\bf k}_{\rm in}, \omega)$ at different input/output angles for SMT and RCM, the confocal reflection coefficient $R_{\rm c}({\bf r}_{\parallel}, \omega)$ at different positions ${\bf r}_{\parallel}$ for OCT, OCM, and ISAM.
To compute these quantities, we need to set up the appropriate source profile [$b_m({\bf r})$ in Eq.~\eqref{eq:wave_eq}] that generates the desired illumination field $E_m^{\rm in}({\bf r})$, compute the resulting total field $E_m^{\rm tot}({\bf r})$, and project the scattered field $E_m^{\rm sca}({\bf r}) = E_m^{\rm tot}({\bf r}) - E_m^{\rm in}({\bf r})$ onto the detectors to model the desired detection schemes.

The illumination field $E_{\rm in}({\bf r})$ is a solution of the homogeneous wave equation [with $\varepsilon_{\rm r}({\bf r})=n_{\rm air}^2$ everywhere].
For $R_{\rm c}({\bf r}_{\parallel}, \omega)$, we let $E_{\rm in}$ be a normal incident Gaussian beam in the air with focal spot at ${\bf r}_{\rm f}^{\rm air} = ({\bf r}_{\parallel}, z_{\rm f}^{\rm air})$.
The apparent focal depth $z_{\rm f}^{\rm air}$ is chosen such that in the presence of the medium, the Gaussian beam will focus to ${\bf r}_{\rm f} = ({\bf r}_{\parallel}, z_{\rm f})$ with the actual focal plane depth at $z_{\rm f}$ (Supplementary Sec.~2).
With OCT, we set the NA of the Gaussian beam to $0.04$ so that its depth of focus covers the full imaging range of interest.
With OCM and ISAM, we set the Gaussian beam NA to 0.5.
The lateral position ${\bf r}_{\parallel}$ is scanned while $z_{\rm f} = 150$ \textmu m is fixed.

For the angular reflection matrix $R({\bf k}_{\rm out}, {\bf k}_{\rm in}, \omega)$, we let $E_{\rm in}({\bf r})$ be a truncated plane wave with a field of view of FOV $=$ 400 \textmu m.
The plane wave is truncated with a Planck--taper window having a full width at half maximum  (FWHM) equaling the FOV centered around $y=0$ at $z = z_{\rm f}^{\rm air}$, which is also the reference plane of the plane wave.
We place the reference plane of the angular reflection matrix at the same depth $z_{\rm f}^{\rm air}$ as the focal plane in $R_{\rm c}({\bf r}_{\parallel}, \omega)$.

Having specified the illumination field $E_{\rm in}({\bf r})$, we use a line source [with $b_m({\bf r})$ being a delta function in $z$ at $z_{\rm source}$] in the air to generate $E_m^{\rm in}({\bf r})$ based on $E_m^{\rm in}(y,z=z_{\rm source})$~\cite{2024_Lin_JphysPhoton}.
We ensure that the simulation domain is large enough that $b_m({\bf r})$ is negligibly small inside the PML.

Given the source profile $b_m({\bf r})$, we can compute the total field profile $E_m^{\rm tot}({\bf r})$ by solving Eq.~\eqref{eq:wave_eq}.
Figure~\ref{fig:simulation_setup}c--f show examples of $|E_m^{\rm tot}({\bf r})|^2$ for a truncated-plane-wave input used in the computation of $R({\bf k}_{\rm out}, {\bf k}_{\rm in}, \omega)$ and for a Gaussian-beam input used in the computation of $R_{\rm c}({\bf r}_{\parallel}, \omega)$.
Because of multiple scattering in the medium, the intensity profile $|E_m^{\rm tot}({\bf r})|^2$ is far more complex than a truncated plane wave or a Gaussian beam.
We also verify that $E_m^{\rm tot}({\bf r})$ agrees with the expected $E_m^{\rm in}({\bf r})$ when the scattering medium is removed.

To model the desired detection schemes, we project the $m$-th scattered field profile onto the $n$-th output of interest as $\int b_n^*({\bf r}) E_m^{\rm sca}({\bf r}) d{\bf r} $,
for both the confocal detection of $R_{\rm c}({\bf r}_{\parallel}, \omega)$ and the plane-wave detection of $R({\bf k}_{\rm out}, {\bf k}_{\rm in}, \omega)$.

\vspace{-6pt}
\subsection{Numerical computation}
\vspace{-4pt}

For the angular reflection matrix $R({\bf k}_{\rm out}, {\bf k}_{\rm in}, \omega)$, we sample the input and output angles with momentum spacing $\delta k_y \approx 2\pi/{\rm FOV}$ within NA = 0.5; this requires scanning between 445 and 635 incident angles per frequency.
For the confocal reflection map $R_{\rm c}({\bf r}_{\parallel}, \omega)$, we sweep the focus of the Gaussian beam across 300 \textmu m in $y$ with 0.2 \textmu m separation (for OCM and ISAM, for which NA = 0.5) or 2 \textmu m separation (for OCT, for which NA = 0.04); this requires scanning 1500 incident Gaussian beams per frequency for OCM and ISAM, 150 for OCT.

Performing full-wave simulations for such a large system over such a large number of inputs would normally require prohibitive computing resources.
We reduce the computing cost by recognizing that it is not necessary to compute the full field profile $E_m^{\rm tot}({\bf r})$ on every pixel of the simulation domain as conventional methods do.
Instead, we directly evaluate the signal at the detectors, namely the projected quantities $\int b_n^*({\bf r}) E_m^{\rm sca}({\bf r}) d{\bf r}$.
The ``augmented partial factorization'' (APF) approach we recently introduced~\cite{Hsu2022} allows us to evaluate those projected quantities while bypassing the full-field solution and to handle
all inputs simultaneously, with no loop over the hundreds of source profiles.
Doing so reduces computing time by orders of magnitude. 

We use the APF method implemented in software MESTI~\cite{MESTI_GitHub} to perform all simulations here.
Computing the angular reflection matrix $R({\bf k}_{\rm out}, {\bf k}_{\rm in}, \omega)$ for all input and output angles or computing the confocal reflection coefficient $R_{\rm c}({\bf r}_{\parallel}, \omega)$ for all positions ${\bf r}_{\parallel}$ takes about 4 minutes per frequency when using one core of Intel Xeon Gold 6130.
The different frequencies can be computed in parallel using different cores of the same node and/or using different computing nodes. 
The memory usage is around 17 GB.

\vspace{-6pt}
\subsection{Image reconstruction}
\vspace{-4pt}

After computing $R({\bf k}_{\rm out}, {\bf k}_{\rm in}, \omega)$ and $R_{\rm c}({\bf r}_{\parallel}, \omega)$ from the simulations, we add complex Gaussian noises with a relative amplitude of 10\% to the data 
(based on the actual noise amplitude in an SMT experiment~\cite{Hsu2023})
and then use the noisy data to reconstruct images of the targets within $|y| < 150$ \textmu m and $ 0 < z < 300$ \textmu m.

\begin{figure*}
    \includegraphics[width=0.85\textwidth]{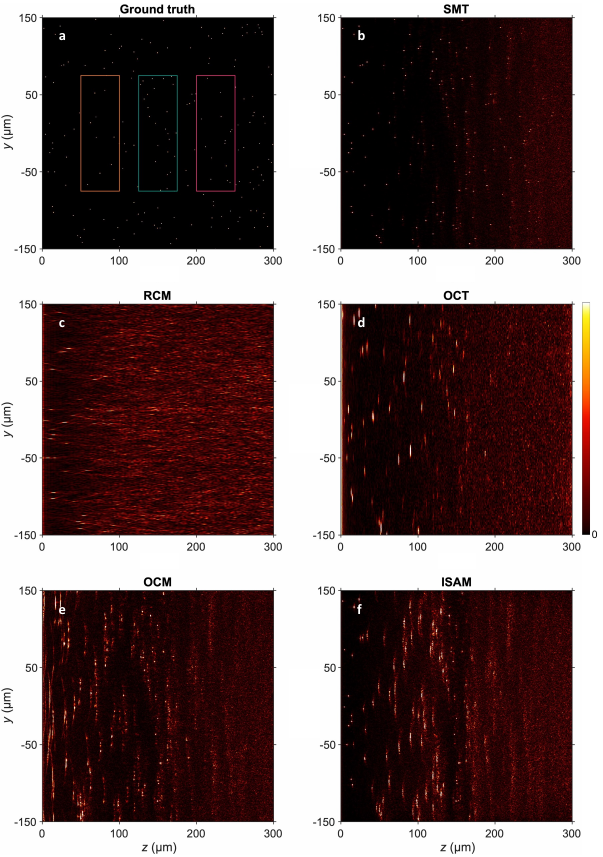}
    \caption{\label{fig:full-view} {\bf Reconstructed images.}
    {\bf a}, Ground-truth locations of the high-index scatterers in Fig.~\ref{fig:simulation_setup}a--b; the low-index ones are not shown. 
    The colored rectangular boxes indicate the zoom-in regions of Fig.~\ref{fig:zoom-in}.
    {\bf b--f}, Images reconstructed from full-wave simulations that model scattering matrix tomography (SMT), reflectance confocal microscopy (RCM), optical coherence tomography (OCT), optical coherence microscopy (OCM), and interferometric synthetic aperture microscopy (ISAM). 
    The focal plane of $R_{\rm c}({\bf r}_{\parallel}, \omega)$ for OCT, OCM, and ISAM and the reference plane of $R({\bf k}_{\rm out}, {\bf k}_{\rm in}, \omega)$ for SMT are both placed at $z_{\rm f} = 150$ \textmu m.
    NA = 0.04 for OCT; NA = 0.5 for all other methods.
    The wavelengths cover $\lambda \in [700, 1000]$ nm.
    }
\end{figure*}

The SMT and RCM reconstructions follow from Eq.~\eqref{eq:psi_smt} and Eq.~\eqref{eq:psi_rcm} after shifting the reference plane of $R({\bf k}_{\rm out}, {\bf k}_{\rm in}, \omega)$ to $z=0$.
For SMT, we use momenta ${\bf k}_{\rm in/out}$ in the medium (with $n_{\rm eff}$ and accounting for numerical dispersion) to perform the digital spatiotemporal focusing, which corrects for the index-mismatch aberration from the air-medium interface.
For RCM, we use data at the center frequency $\omega_{\rm RCM} = (\omega_{\rm min} + \omega_{\rm max})/2$ and synthesize confocal foci in the air and then convert the apparent depth to the actual focal depth in the medium (Supplementary Sec.~2).
We perform the summations in Eq.~\eqref{eq:psi_smt} and Eq.~\eqref{eq:psi_rcm} through non-uniform fast Fourier transforms (NUFFTs) using the FINUFFT library~\cite{Barnett2019}. 

Eq.~\eqref{eq:psi_ocm_or_oct} for OCT and OCM neglects the index mismatch at the air-medium interface.
Since the reference plane of $R_{\rm c}({\bf r}_{\parallel}, \omega)$ has confocal illumination and detection at $z_{\rm f}^{\rm air}$ in the air, we need to include an additional delay in $\Delta t$ to account for the additional distance from $z_{\rm f}^{\rm air}$ to $z_{\rm f}$ and the difference in propagation speed in air and in the medium.
Additionally, numerical dispersion gives rise to a frequency-dependent group velocity.
To include the additional time delay and to correct for the numerical group velocity dispersion, we replace the $-2\omega({z - z_{\rm f}})/{v_{\rm g}}$ phase factor in Eq.~\eqref{eq:psi_ocm_or_oct} with $-2(k_{{\rm eff}}z-k_{{\rm air}}z_{\rm f}^{\rm air})$ where $k_{\rm eff}$ and $k_{\rm air}$ are the wave numbers at normal incidence in the medium (with index $n_{\rm eff}$) and in the air respectively. 
The $-k_{{\rm air}}z_{\rm f}^{\rm air}$ term shifts the reference plane to the air-medium interface, and the $k_{{\rm eff}}z$ term accounts for the propagation phase shift to depth $z$.

The ISAM image uses the same $R_{\rm c}({\bf r}_{\parallel}, \omega)$ data as OCM. The reconstruction procedure follows the description in Sec.~\ref{sec:imaging_methods}, with the additional details that we shift the reference plane of $R_{\rm c}({\bf r}_{\parallel}, \omega)$ to $z=0$, that we account for numerical dispersion when computing $q_z = 2 k_z^{\rm out}$, and that we use NUFFT to evaluate Eq.~\eqref{eq:psi_isam} so it is not necessary to interpolate ${\bf q}$ onto a rectangular grid.

Each of these (300 \textmu m)$^2$ images takes less than one minute to build on a MacBook Air with M1 chip.

In the raw images, the peak intensity varies by many orders of magnitude across the 300 \textmu m range in depth. There is an exponential decay of the signal due to multiple scattering and a slow decay of the diffusive background. For OCM and ISAM, there is also an increased intensity near the focal plane $z_{\rm f} = 150$ \textmu m.
These variations make it hard to directly visualize the raw images across the full volume.
Therefore, we normalize each of the images with a factor that slowly varies with $z$ to reduce the variation of the peak intensity (Supplementary Sec.~3).

To validate our image reconstruction algorithms and our setup for numerically computing $R({\bf k}_{\rm out}, {\bf k}_{\rm in}, \omega)$ and $R_{\rm c}({\bf r}_{\parallel}, \omega)$, we show in Supplementary Sec.~4 the reconstructed images of a weakly scattering system where we remove all of the low-index-contrast scatterers.
This system is in the single-scattering regime since its scattering mean free path $l_{\rm sca} \approx 390$ \textmu m (Supplementary Sec.~1) is larger than the imaging depth.
In this single-scattering system, SMT accurately captures all of the 194 high-index targets within the (300 \textmu m)$^2$ area, and the other methods capture nearly all (over 95\%) of them, failing only for clusters of close-by targets.
As expected, the lateral resolution of OCM degrades severely away from the focal plane, which is restored perfectly by ISAM across all depths.
The locations of all of the targets agree quantitatively with the ground truth.

In Ref.~\cite{Hsu2023}, the reflection matrix $R({\bf k}_{\rm out}, {\bf k}_{\rm in}, \omega)$ was also used to synthesize OCT and OCM images.
In Supplementary Sec.~5, we adopt the same scheme and verify that such synthetic OCT/OCM images indeed reproduce the OCT/OCM images built from $R_{\rm c}({\bf r}_{\parallel}, \omega)$.

\begin{figure*}
    \includegraphics[width=0.9\textwidth]{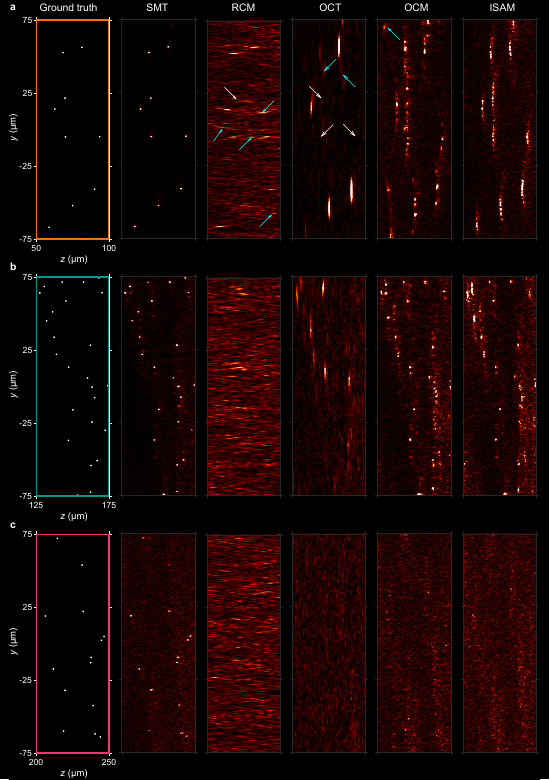}
    \caption{\label{fig:zoom-in}{\bf Zoom-in views of the reconstructed images.}
    Each row shows a zoom-in of the images in Fig.~\ref{fig:full-view}.
    Cyan and white arrows in {\bf a} indicate false-positive artifacts and false-negative missing targets respectively identified through comparison to the ground truth. All panels use the same colorbar and the same scales as in Fig.~\ref{fig:full-view}.
    }
\end{figure*} 

\begin{figure*}
    \includegraphics[width=0.95\textwidth]{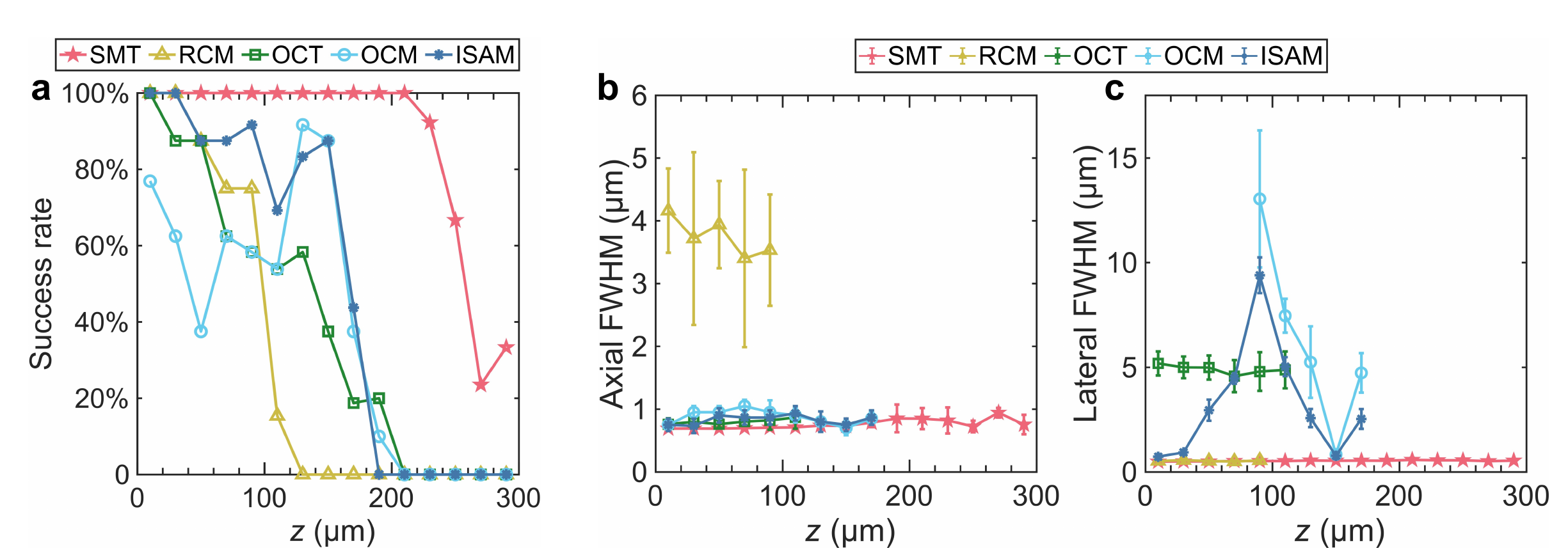}
    \caption{\label{fig:success-rate} {\bf Performance metrics of the imaging methods.} 
    {\bf a}, Success rates in identifying the high-index scatterers based on their ground-truth positions. 
    {\bf b-c}, The axial and lateral FWHM resolutions as a function of depth. 
    Error bar length is twice the standard deviation.
    Data are shown only for depths where at least 4 targets were successfully identified.
    The lateral resolution of OCM in the front is not shown as the defocusing is too severe for a FWHM to be defined there.
    }
\end{figure*}

\vspace{-4pt}
\section{Imaging Results}
\vspace{-2pt}

We now examine images reconstructed from full-wave simulations on the strongly scattering system described in Sec.~\ref{sec:model_system} and Fig.~\ref{fig:simulation_setup}a--b. 
Figure~\ref{fig:full-view} shows the ground-truth locations of the high-index target scatterers (enlarged slightly for ease of viewing) and the reconstructed SMT, RCM, OCT, OCM, and ISAM images. 
For a clearer view, Fig.~\ref{fig:zoom-in} shows close-ups of these images near the front of the sample (Fig.~\ref{fig:zoom-in}a), near the focal/reference plane (Fig.~\ref{fig:zoom-in}b), and deeper inside (Fig.~\ref{fig:zoom-in}c).  

To quantify the imaging performance, we zoom in further onto each of the target scatterers based on their ground-truth positions.
We mark each target as successfully identified if the reconstructed image has a single sharp peak within plus/minus one axial and one lateral FWHM resolution from the ground-truth location of that target. If there are multiple bright spots with comparable intensities near the ground-truth location, we consider those bright spots as speckles coming from multiple scattering that overwhelm the target signal, so the target identification is marked as unsuccessful.
The success rates are plotted in Fig.~\ref{fig:success-rate}a as a function of depth. 

SMT achieves a perfect success rate down to around 250~\textmu m depth, beyond which it sharply begins to fail as the intensity of the multiple-scattering background overtakes the target signals.
It also attains the highest signal-to-background ratio among all methods. 
This is thanks to the digital confocal spatiotemporal gating and index-mismatch correction that restores the sharpness and brightness of the spatial focus (Supplementary Secs.~2 and 6).

The other images exhibit artifacts even at relatively shallow depths where these methods are expected to work.
For the front of the sample in Fig.~\ref{fig:zoom-in}a, we mark false-negative missed targets with white arrows (which lower the success rate above) and false-positive artifacts with cyan arrows (which are not considered in the success rate).
These artifacts arise from multiple scattering.
Without knowledge of the ground truth (which is typically the case in experiments), artifacts like these would have gone unnoticed, and we would have falsely interpreted these images as faithful representations of the sample at these depths.

We then measure the axial and lateral FWHM resolutions for the successfully identified targets, with the mean and standard deviation shown in Fig.~\ref{fig:success-rate}b--c.
The axial resolution of RCM is low due to the lack of time gating, which also led to the worst signal-to-background ratio.
While OCT provides a good signal-to-background ratio at shallow depths, it has a poor lateral resolution due to the small NA and minimal spatial gating. 
OCM achieves good lateral and axial resolutions near the focal plane, but its lateral resolution degrades quickly away from the focal plane due to defocusing.
ISAM restores the lateral resolution of OCM near the front surface of the sample at depths $z \lesssim 30$ \textmu m (on the order of $l_{\rm sca}$), but the lateral resolution is still poor at other depths because its synthetic aperture does not provide the filtering of a confocal gate away from the focal plane.
SMT achieves sub-micron resolutions in both axial and lateral directions at all depths.

Above, the focal plane of $R_{\rm c}({\bf r}_{\parallel}, \omega)$ for OCT, OCM, and ISAM and the reference plane of $R({\bf k}_{\rm out}, {\bf k}_{\rm in}, \omega)$ for SMT are both placed at $z_{\rm f} = 150$ \textmu m.
Supplementary Sec.~7 shows OCM and ISAM images with a deeper focal plane at $z_{\rm f} = 250$ \textmu m, though doing so does not improve the success rate.

\vspace{-6pt}
\section{Discussion}
\vspace{-3pt}

The full-wave simulations developed in this work provide 
unique capabilities that complement experimental studies.
Knowledge of the ground-truth configuration allows rigorous assessments of the reconstructed images, leading to the identification of artifacts that would otherwise be mistaken as being correct.
The ability to implement different source and detection schemes in the same virtual setup simplifies and improves the comparison of imaging methods.
One can freely tailor the refractive index profile of the system, making it possible to systematically study the role of individual variables such as the size or the spatial correlations of the scatterers.
The numerical simulations also provide an inexpensive and convenient sandbox for exploring new imaging methods and algorithms.
We have recently implemented the APF method for 3D vectorial waves~\cite{MESTI_3D_GitHub}.
Future work may adopt full-wave modeling in 3D as well as tissue phantom models~\cite{Schmitt1998, Lu2005}.

\vspace{12pt}
\noindent
{\bf Acknowledgments:} We thank H.-C.~Lin, B.~Applegate, A.~Goetschy, M.~Dinh, and X.~Gao for helpful discussions. 
{\bf Funding:} This work was supported by the Chan Zuckerberg Initiative and the National Science Foundation (ECCS-2146021).
Computing resources are provided by the Center for Advanced Research Computing at the University of Southern California.
{\bf Competing interests:} The authors declare that they have no competing interests.
{\bf Data and materials availability:} All data needed to evaluate the conclusions in the paper are present in the paper and the Supplementary Materials.
Simulation and image reconstruction codes are available on GitHub~\cite{GitHub_imaging_simulations}.

\bibliography{main}

\end{document}


\title{Supplementary Materials\\
Full-wave simulations of tomographic optical imaging inside scattering media
}
\author{Zeyu Wang}
\author{Yiwen Zhang}
\author{Chia Wei Hsu}
\affiliation{%
Ming Hsieh Department of Electrical and Computer Engineering, University of Southern California, Los Angeles, California 90089, USA
}%

\maketitle

\tableofcontents

\section{Mean free path estimation} 
In this section, we estimate the scattering mean free path $l_{\rm sca}$, transport mean free path $l_{\rm tr}$, and anisotropy factor $g$ in our model system, accounting for the effect of spatial correlations in the positions of the scatterers. 

First, we compute the differential scattering cross section ${d\sigma_{m}}/{d\Omega}$ for the two types of cylinders ($m=1,2$) in the system. In 2D, the solid angle $\Omega$ equals the scattering angle $\theta$.
For an accurate estimation, we numerically compute ${d\sigma_{m}}/{d\Omega}$ by considering the same finite-difference wave equation with grid size $\Delta x=0.1$ \textmu m as the model system, solving for the scattered wave from a single cylinder in a background with refractive index $n_{\rm bg} = 1.33$ given a plane-wave input, and performing near-field-to-far-field transformation using the far-field Green's function of the finite-difference wave equation~\cite{Martin2006}.
Due to discretization and subpixel smoothing, the cylinder's permittivity profile depends on the cylinder's center location with respect to the grid.
We average the result over 100 random center locations. 

Given the large filling fraction (around 10\%) of the low-index-contrast cylinders, the independent-particle approximation~\cite{Carminati2021} is not accurate, and we need to account for the correlations in their positions. We estimate $l_{\rm sca}$ by~\cite{Kevin2021}
\begin{equation}
    \frac{1}{l_{\rm sca}} = \sum_{m} \rho_{m} \int \frac{d\sigma_{m}}{d\Omega}(\theta) S_{m}(\theta) d\Omega
    \label{eq:l_sca_eq},
\end{equation}
where $\rho_m$ is the number density of scatterer type $m$, and $S_{m}(\theta)$ is its structure factor. Similarly, we estimate $l_{\rm tr}$ by
\begin{equation}
    \frac{1}{l_{\rm tr}} = \sum_{m} \rho_{m} \int \frac{d\sigma_{m}}{d\Omega}(\theta)S_{m}(\theta)(1-\cos\theta) d\Omega.
    \label{eq:l_tr_eq}
\end{equation}
The anisotropy factor $g$ is then given by $g = 1 - l_{\rm sca}/l_{\rm tr}$, and the scattering cross section is $\sigma_m = \int \frac{d\sigma_{m}}{d\Omega} d\Omega$. 
We compute the structure factor from the scatterer positions by~\cite{McDonald2013}
\begin{equation}
    S_m(|{\bf q}|) = \frac{1}{N_m} \bigg\langle \bigg|\sum_{j=1}^{N_m} e^{i{\bf q}\cdot {\bf r}_j^{(m)}}\bigg|^2 \bigg \rangle,
    \label{eq:structure_factor}
\end{equation}
where $N_m$ is the number of scatterers of type $m$, ${\bf r}_j^{(m)}$ is the $j$-th scatterer location, ${\bf q} = {\bf k}_{\rm out} - {\bf k}_{\rm in}$ is the momentum transfer, and $\langle \cdots \rangle$ averages over different directions of ${\bf q}$ with the same $|{\bf q}|$. We smooth $S_m(|{\bf q}|)$ with a spline and then convert the momentum transfer to the scattering angle $\theta$ through $|{\bf q}|=2k\sin(\theta/2)$, where $k$ is the wave vector in the background medium. 

Table~\ref{tab:scatterer-parameters} summarizes the scattering parameters at the lowest and highest wavelengths and averaged over wavelengths.

\begin{table*}[ht]
\begin{tblr}{ 
    width = 0.6\textwidth, 
    hlines, vlines, 
    rows = {halign=c, valign=m}, 
    column{1} = {1.5cm}, 
    column{2} = {1.2cm},
    column{3} = {0.6cm},
    column{4} = {1.0cm},
    cell{2}{1,2,3,4} = {r=3}{c},
    cell{5}{1,2,3,4} = {r=3}{c},
    } 
        & Diameter (\textmu m) & $\varepsilon$ & $\rho_m$ (1/\textmu m$^2$) & $\lambda$ (\textmu m) & $\sigma_m$ (\textmu m) & $l_{\rm sca}$ (\textmu m) & $l_{\rm tr}$ (\textmu m) & $g$ \\ 
        low-index-contrast cylinders & 0.7 & $1.4^2$ & 0.25 & 0.7 & 0.14 & 35 & 404 & 0.91 \\
         & & & & 1.0 & 0.055  & 91 & 597  & 0.85 \\
         & & & & 0.7--1.0 & 0.088 &  61 & 538 & 0.89 \\
        high-index cylinders \ (targets) & 0.3 & $2.5^2$ & 0.002 & 0.7 & 0.80 & 642 & 864 & 0.26\\
         & & & & 1.0 & 1.37 & 372 & 1090 & 0.66\\
         & & & & 0.7--1.0 & 1.35 & 391 & 1010 & 0.60\\
        total &  &  &  & 0.7--1.0 &  & 53 & 351 & 0.85
\end{tblr}
\caption{\label{tab:scatterer-parameters} \bf Scattering parameters of the model system.}
\end{table*}

\section{Apparent depth and aberration from index mismatch }\label{sec:refraction}

\begin{figure*}
    \includegraphics[width=1.0\textwidth]{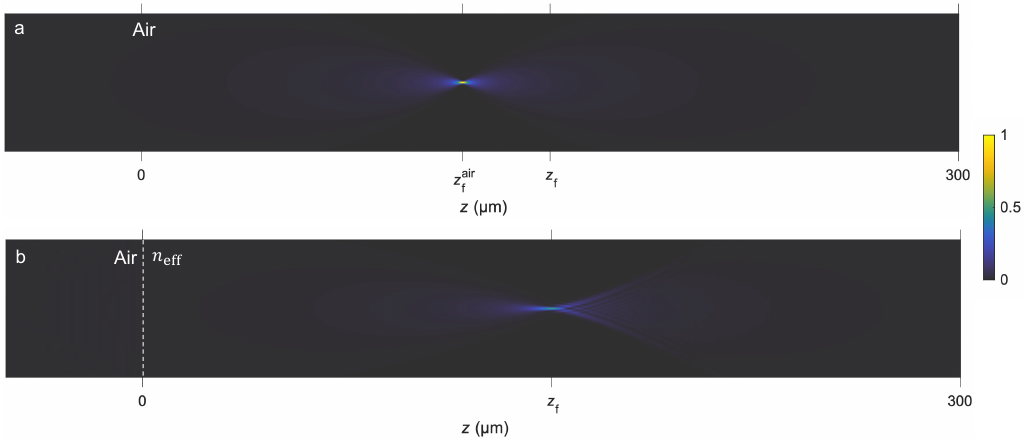}
    \caption{\label{fig:focus_profile} {\bf Effects of refraction on an incident beam.} 
    Intensity profiles of the same incident Gaussian beam in air (NA = 0.5) as in Fig.~1e--f of the main text but with $\varepsilon({\bf r})$ being air everywhere ({\bf a}) and when there is an air-medium interface at $z=0$ but no scatterers ({\bf b}).
    Both are computed from full-wave simulations for the discretized wave equation with $\Delta x = 0.1$ \textmu m at wavelength $\lambda = 823$ nm.
    The desired focal depth in the medium is $z_{\rm f}=150$ \textmu m, and Eq.~\eqref{eq:depth_scaling_factor} is used to find the apparent focal depth $z_{\rm f}^{\rm air} = z_{\rm f}/\alpha \approx 118$ \textmu m in air for the incident beam.    
     }
\end{figure*}

When modeling OCT, OCM, and ISAM in our system, we consider illumination and detection with Gaussian beams in air ({\bf Fig.}~\ref{fig:focus_profile}{\bf a}) having an apparent focal depth of $z_{\rm f}^{\rm air}$.
At the air-medium interface, light refracts, so the Gaussian beam focuses to a deeper depth $z_{\rm f}$ and elongates along the axial direction ({\bf Fig.}~\ref{fig:focus_profile}{\bf b}).
It is well known that 
$z_{\rm f} / z_{\rm f}^{\rm air} \approx n_{\rm medium}/n_{\rm air}$
at small angles.
This section derives the angle-dependent apparent depth while accounting for numerical dispersion from discretization.

Consider the interface between two homogeneous media with
\begin{equation}
n(z) = 
\begin{cases}
  n_1, \quad z \le 0, \\    
  n_2, \quad z > 0.
\end{cases}
\end{equation}
Ignoring the Fresnel reflection/transmission coefficients (which is not important here), the TM fields in 2D satisfy
\begin{equation}
\label{eq:E_interface}
E_x(y,z) = \sum_{k_y} e^{ik_y(y-y_{\rm f})}
\begin{cases}
  e^{ik_{z, 1}(z-z_{{\rm f}, 1})}, \qquad z \le 0,\\    
  e^{ik_{z, 2}z-ik_{z, 1}z_{{\rm f}, 1}}, \quad z > 0.
\end{cases}
\end{equation}
Here, $(y_{\rm f}, z_{{\rm f}, 1})$ is the apparent focal spot:
we can see that all phases in the exponential line up at $(y,z) = (y_{\rm f}, z_{{\rm f}, 1})$ if we extrapolate the $z \le 0$ expression to $z > 0$.
Here, $k_{z, j}$ with $j=1,2$ is the longitudinal wave number in medium $j$;
in the discretized wave equation, it is related to the transverse wave number $k_y$ (which is invariant given the continuous translational symmetry in $y$) through the finite-difference dispersion relation~\cite{Hsu2022}
\begin{equation}
\label{eq:FDFD_dispersion}
    \left(\frac{\omega}{c} n_j \Delta x\right)^2 = 4\sin^2\left(\frac{k_y\Delta x}{2}\right) + 4\sin^2\left(\frac{k_{z,j}\Delta x}{2}\right),
\end{equation}
which is derived by inserting a plane wave profile $e^{ik_y y + i k_{z, j}z}$ into the discretized wave equation with a homogeneous refractive index $n_j$. Eq.~\eqref{eq:FDFD_dispersion} reduces to the continuous dispersion relation $(\omega n_j/c)^2 = k_y^2 + k_{z,j}^2$ when $\Delta x \to 0$. We want to determine the actual focal spot $(y_{\rm f}, z_{{\rm f}, 2})$ in the $n_2$ medium, which is where the phases in the exponential of Eq.~\eqref{eq:E_interface} line up.
For a given angle with transverse momentum $k_y$, the nearby angles have their phases lined up when the phase is independent of $k_y$ to leading order:
\begin{equation}
\frac{d}{dk_y}(k_{z, 2}z_{{\rm f}, 2}-k_{z, 1} z_{{\rm f}, 1}) = 0, 
\end{equation}
which yields, through Eq.~\eqref{eq:FDFD_dispersion},
\begin{equation}
\label{eq:z_ratio}
  \frac{z_{{\rm f}, 2}(k_y)}{z_{{\rm f}, 1}}
  = \frac{dk_{z, 1}/dk_y}{dk_{z, 2}/dk_y}  
  = \frac{\sin(k_{z, 2}\Delta x)}{\sin(k_{z, 1}\Delta x)}.
\end{equation}
Note that this ratio depends on the incident angle.
In the continuous limit $\Delta x \rightarrow 0$, Eq.~\eqref{eq:z_ratio} reduces to 
$z_{{\rm f}, 1}\tan(\theta_1) = z_{{\rm f}, 2}\tan(\theta_2)$,
consistent with geometrical optics.
The angle dependence means that given a fixed apparent focal depth $z_{{\rm f}, 1}$, different angular components of the Gaussian beam will focus to different focal depths $z_{{\rm f}, 2}(k_y)$, which stretches the axial extent of the Gaussian beam and lowers its peak intensity, forming spherical aberrations.
The axial profile of the beam is asymmetrical, and its center is not well defined.
We define the beam center $(y_{\rm f}, z_{{\rm f}, 2})$ through
\begin{equation}
  \frac{z_{{\rm f}, 2}}{z_{{\rm f}, 1}}
  = \left\langle \frac{\sin(k_{z, 2}\Delta x)}{\sin(k_{z, 1}\Delta x)} \right\rangle_{k_y}
  \equiv \alpha,
\label{eq:depth_scaling_factor}
\end{equation}
where $\langle \cdots \rangle_{k_y}$ denotes averaging over the transverse momentum $k_y$ within the NA of the Gaussian beam.
For small angles in the continuous limit $\Delta x \to 0$, $\alpha \approx n_2/n_1$.

For a discretized system with grid size $\Delta x = 0.1$ \textmu m at wavelength $\lambda = 823$ nm, $n_1 = 1$, $n_2 = n_{\rm eff} \approx 1.34$, and angular range NA = 0.5, Eq.~\eqref{eq:depth_scaling_factor} yields $\alpha \approx 1.27$.
Figure~\ref{fig:focus_profile} shows the intensity profile of a Gaussian beam in air with NA = 0.5, computed from full-wave simulations following the procedure described in Sec.~II C of the main text.
We set the desired focal depth at $z_{{\rm f}, 2} = z_{\rm f} = 150$ \textmu m, compute the apparent focal depth $z_{{\rm f}, 1} = z_{\rm f}^{\rm air} = z_{\rm f}/\alpha \approx 118$ \textmu m in air from Eq.~\eqref{eq:depth_scaling_factor}, put the line source in the air on the left, and compute the field profile either with air everywhere ({\bf Fig.}~\ref{fig:focus_profile}{\bf a}) or with air on the left and a homogeneous medium with index $n_{\rm eff} \approx 1.34$ on the right.
In both cases, the Gaussian beam focuses to the expected location.
In addition to the focus shift, we can also observe how refraction at the air-medium interface elongates the focus and reduces its peak intensity.
Note that numerical discretization reduces the conversion factor $\alpha$, and the depth would be inaccurate if we used the continuous expression for $\alpha$.

\section{Depth-dependent image intensity normalization}

\begin{figure*}
    \includegraphics[width=1.0\textwidth]{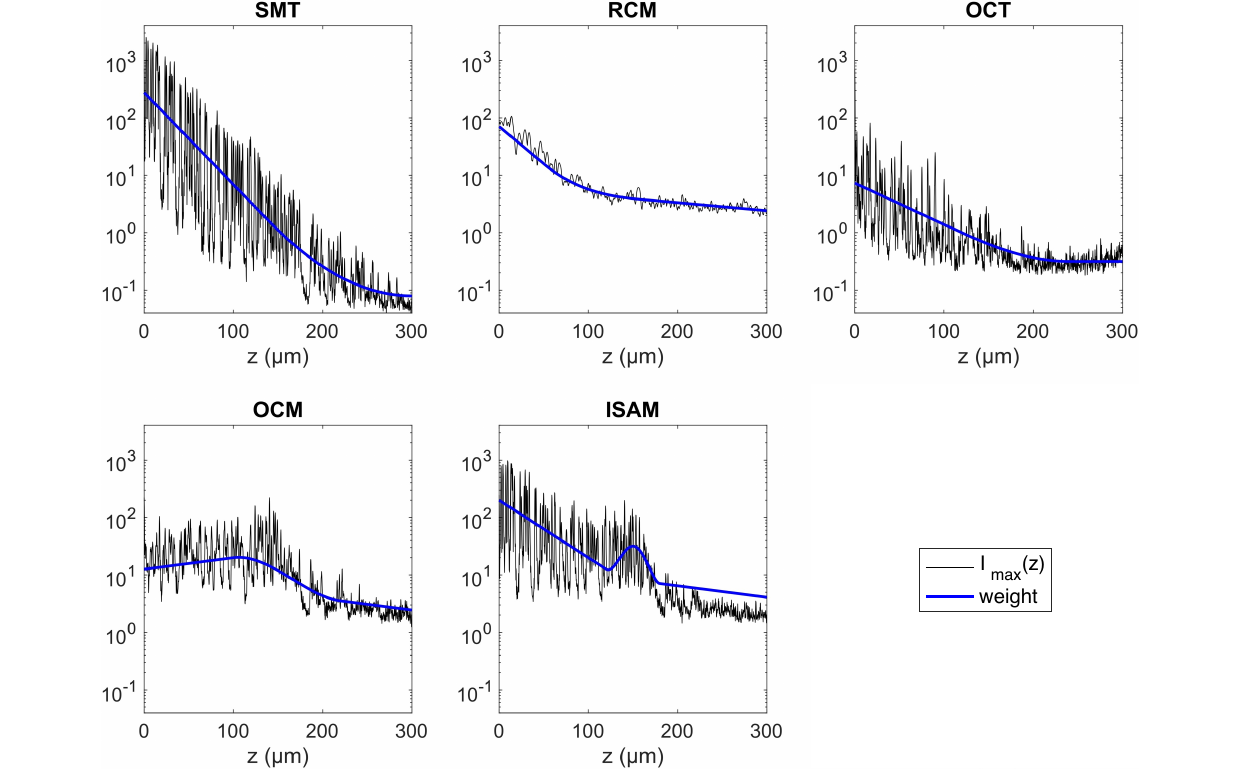}
    \caption{\label{fig:z_dependent_weights} {\bf Depth-dependent peak intensity.} 
    The black lines show the peak intensity at each depth, $I_{\rm max}(z) \equiv {\rm max}_{y}\,I(y, z)$, for each of the reconstructed images for the full strongly scattering system.
    We find smooth curves (blue lines) that capture the overall trend of the peak intensity and normalize the images by these smooth curves before subsequent plotting.
    }
\end{figure*}

In the raw images, the peak intensity varies by many orders of magnitude across the 300 \textmu m range in depth, making direct visualization difficult.
Therefore, we normalize each of the images with a weight that slowly varies with $z$ to reduce the variation of the peak intensity.
To find the weight to use, we plot the peak image intensity at each depth, $I_{\rm max}(z) \equiv {\rm max}_{y}\,I(y, z)$, find a smooth curve that captures the overall trend of $I_{\rm max}(z)$ as shown in {\bf Fig.}~\ref{fig:z_dependent_weights}, and divide the image $I(y,z)$ by this weight before subsequent plotting.
Note that finding such weight does not require the ground truth or the scattering parameters of the system; we only use the reconstructed images to determine the weight.

\section{Validation in a weakly scattering system}

\begin{figure*}
    \includegraphics[width=0.9\textwidth]{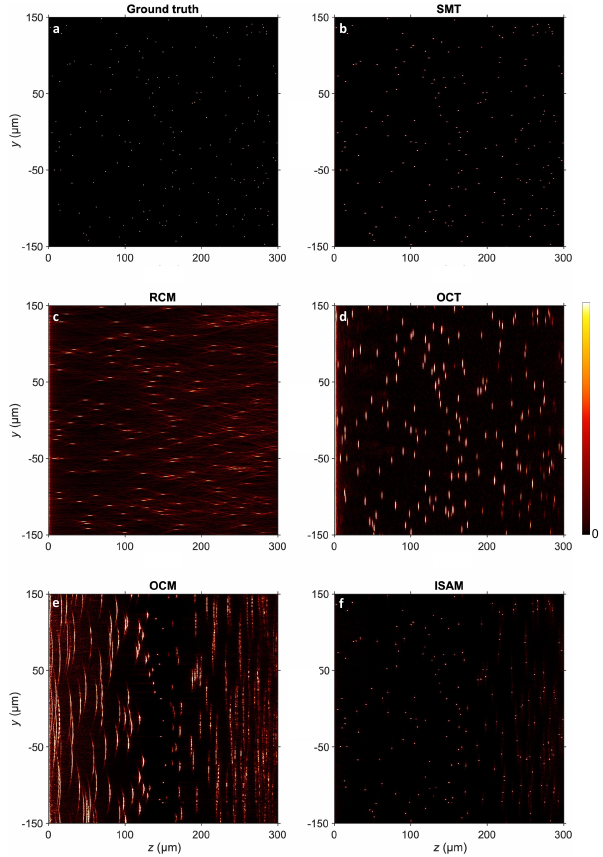}
    \caption{\label{fig:weak-scattering} {\bf Reconstructed images in a weakly scattering system.}
    The low-index-contrast scatterers are removed to reduce the scattering strength.
    The agreement with the ground truth here validates that the simulations and the reconstructions are set up correctly.
    }
\end{figure*}

To verify that the illumination and detection are set up correctly in the simulations and that the image reconstruction methods are implemented correctly, we consider a weakly scattering system where we remove all of the low-index-contrast scatterers, keeping only the high-index targets.
The scattering mean free path $l_{\rm sca} = 391$ \textmu m in such case (Table~\ref{tab:scatterer-parameters}) is larger than the depths we consider, so scattering effects are small.
Figure~\ref{fig:weak-scattering} shows the reconstructed images of the five methods in comparison to the ground truth.
All reconstructions perform as expected here, and the locations of all of the targets agree quantitatively with the ground truth. 

\section{Comparison between synthetic and physical OCT and OCM}

\begin{figure*}
    \includegraphics[width=1.0\textwidth]{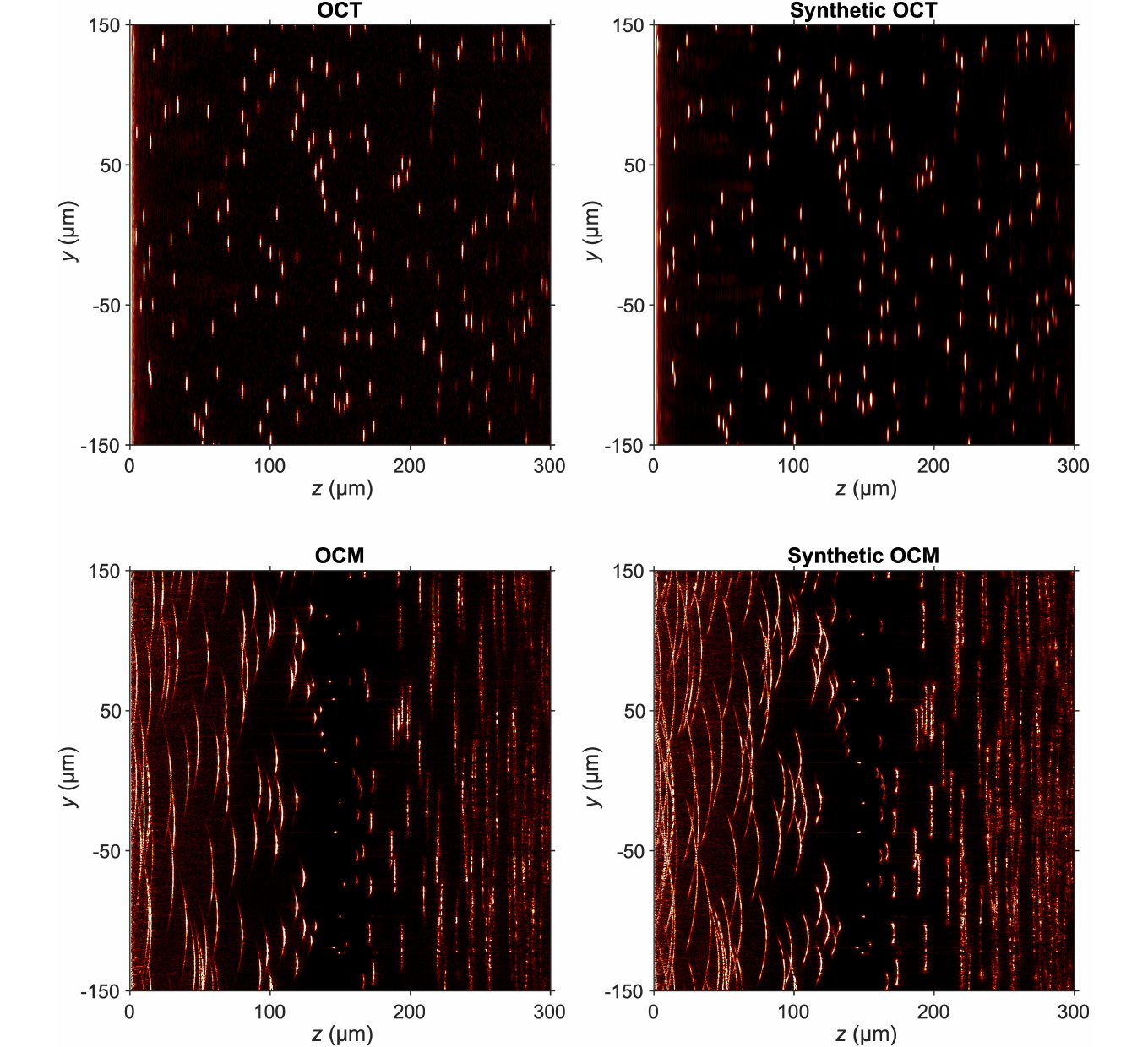}
    \caption{\label{fig:no-weak} {\bf Comparison between synthetic and physical OCT/OCM.}
    The system is the same weakly scattering one as in {\bf Fig.}~\ref{fig:weak-scattering}.
    }
\end{figure*}

Ref.~\cite{Hsu2023} uses the reflection matrix $R({\bf k}_{\rm out}, {\bf k}_{\rm in}, \omega)$ to synthesize OCT and OCM images.
To verify that such synthetic OCT/OCM produces comparable images as the physical OCT/OCM built from the confocal reflection coefficient $R_{\rm c}({\bf r}_{\parallel}, \omega)$,
we show in {\bf Fig.}~\ref{fig:no-weak} such synthetic images (following Supplementary Sec.~IV\,B of Ref.~\cite{Hsu2023}) in comparison to the physical ones, for the same weakly scattering system as considered in the previous section.
The two sets are almost identical.
The minor differences arise because the physical OCT/OCM images here adopt Gaussian beams for illumination and detection, while the synthetic OCT/OCM images sum over plane waves within the NA with a flat non-Gaussian weight (which produces beams with a sinc-function profile in $y$ at the focal plane).

\bigskip

\section{Effects of index-mismatch correction in SMT}

\begin{figure*}
    \includegraphics[width=1.0\textwidth]{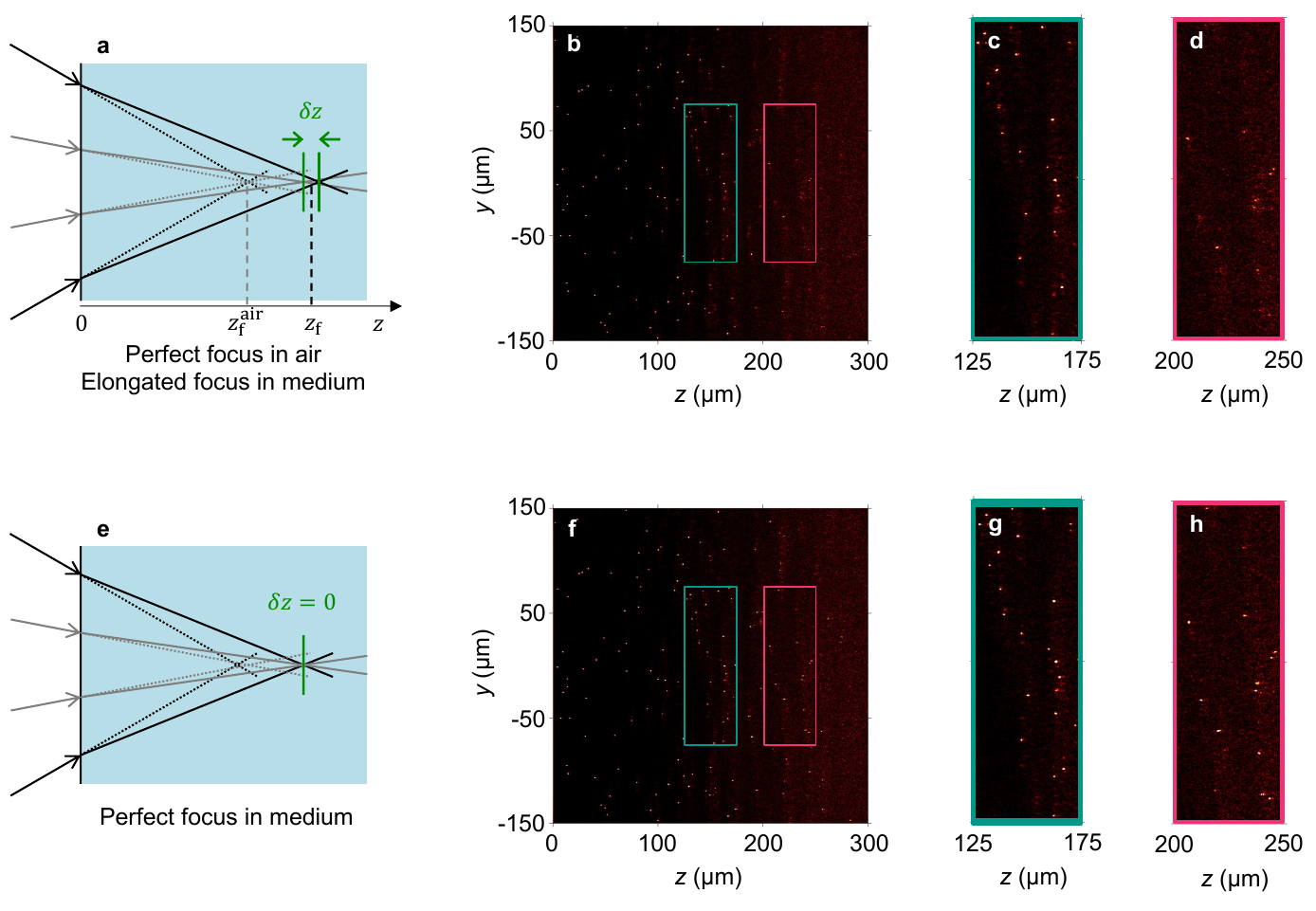}
    \caption{\label{fig:smt_air} {\bf Effects of index-mismatch correction in SMT.} 
    {\bf a} Schematic of an incident beam focused in air (dashed lines), which elongates in the presence of refraction at the air-medium interface because different angular components of the beam converge to different spots (solid lines), reducing the focusing strength. 
    {\bf b-d} Full view and zoom-ins of SMT images of the full strongly scattering system, using such in-air focus without correcting for the index mismatch. 
    {\bf e} Schematic of an incident beam designed to focus in the medium in the presence of refraction. 
    {\bf f-h} Full view and zoom-ins of SMT images of the same system using such in-medium focus, correcting for the index mismatch (same image as in Figs.~2--3 of the main text).
    }
\end{figure*}

In SMT, we digitally correct the aberration created by the refractive index mismatch between interfaces by using the wave vectors $\bf k_{\rm out}$ and $\bf k_{\rm in}$ in the medium in Eq.~(1) of the main text and setting the appropriate reference plane. The details are described in Supplementary Sec.~III\,C of Ref.~\cite{Hsu2023}. {\bf Fig.}~\ref{fig:smt_air} compares SMT images with and without such index-mismatch correction.

There are three differences when we do not perform such index-mismatch correction:
(1) In Eq.~(1) of the main text, we use the wave vectors $\bf k_{\rm out}$ and $\bf k_{\rm in}$ in air instead, so that we synthesize confocal inputs and outputs in air.
(2) We need to replace the actual depth $z = z_{\rm f}$ in the medium with the apparent depth $z' = z_{\rm f}^{\rm air}$ of the input/output focus in Eq.~(1). Here, $z_{\rm f} = \alpha z_{\rm f}^{\rm air}$, as described in Sec.~\ref{sec:refraction} above.
(3) We need to include an additional time delay to align the temporal gate with the spatial gate in the absence of index mismatch.

Without the correction, the incident pulse would focus perfectly in the absence of the medium (dotted lines in {\bf Fig.}~\ref{fig:smt_air}{\bf a}) to a depth $z_{\rm f}^{\rm air}$ at time $t=0$. It takes time $z_{\rm f}^{\rm air}/v_{\rm g}^{\rm air}$ for the pulse to travel from $z=0$ to $z=z_{\rm f}^{\rm air}$ in the absence of the medium (where $v_{\rm g}^{\rm air}$ is the group velocity in air), so the pulse is at the air-medium interface $z=0$ at time $t=-z_{\rm f}^{\rm air}/v_{\rm g}^{\rm air}$. 
In the presence of the medium, refraction at the interface shifts the focus to a deeper depth $z_{\rm f} = \alpha z_{\rm f}^{\rm air}$.
The different angles converge to slightly different spots (solid lines in {\bf Fig.}~\ref{fig:smt_air}{\bf a}) with a difference of $\delta z$, elongating the focal spot.
It takes time $z_{\rm f}/v_{\rm g}$ for the pulse to travel from the air-medium interface $z=0$ to the actual focal spot $z=z_{\rm f}$ in the presence of the medium (where $v_{\rm g}$ is the group velocity in the medium).
So, the pulse arrives at the elongated focus $z_{\rm f}$ at a later time $t = t_{\rm f}= z_{\rm f}/v_{\rm g} -z_{\rm f}^{\rm air}/v_{\rm g}^{\rm air}$ instead of at $t=0$. To align the temporal gate with the spatial gate, we include an additional time-gating factor $e^{-2i\omega t_{\rm f}}$ to Eq.~(1) in the main text. 

To account for the numerical dispersion of the group velocity, we set the time-gating factor $e^{-2i\omega t_{\rm f}} = e^{-2i\omega(z_{\rm f}/v_{\rm g} -z_{\rm f}^{\rm air}/v_{\rm g}^{\rm air})}$ to $e^{-2i(k_{z}^{\rm eff} z_{\rm f}-k_{z}^{\rm air}z_{\rm f}^{\rm air})} = e^{-2i(k_{z}^{\rm eff} z -k_{z}^{\rm air} z/\alpha)}$, where $k_{z}^{\rm eff}$ and $k_{z}^{\rm air}$ are the longitudinal components of the wave vectors at normal incidence in the medium and in the air respectively. 

The index mismatch elongates the focus and lowers its peak intensity. 
This lowers the signal-to-background ratio of the resulting image, so the image quality is reduced at deeper depths ({\bf Fig.}~\ref{fig:smt_air}{\bf b--d}).
Correcting for the index mismatch makes the different angles coverage to the same spot in the presence of refraction at the air-medium refraction ({\bf Fig.}~\ref{fig:smt_air}{\bf e}), which improves the signal-to-background ratio and the image quality ({\bf Fig.}~\ref{fig:smt_air}{\bf f--h}).

\section{OCM and ISAM with a deeper focal plane}

\begin{figure*}
    \includegraphics[width=1.0\textwidth]{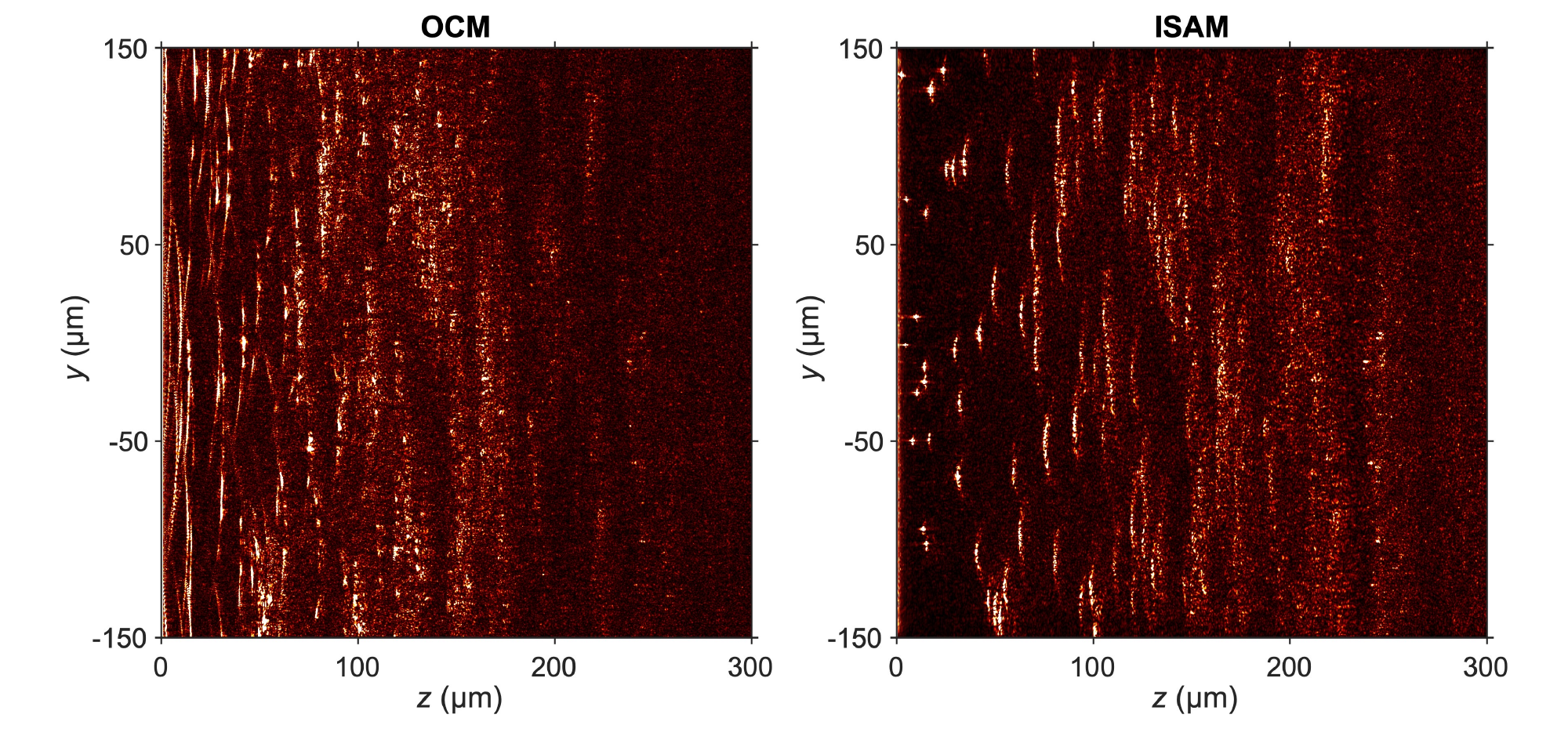}
    \caption{\label{fig:ocm-250-focal-plane} {\bf OCM and ISAM images with the focal plane at $z=250$ \textmu m.} 
    The system is the full strongly scattering one as considered in the main text.
}
\end{figure*}

The OCM and ISAM images depend sensitively on the location of the focal plane $z_{\rm f}$. 
In the main text, we place the focal plane at $z_{\rm f}=150$ \textmu m, same as the reference plane of the reflection matrix used in SMT.
{\bf Fig.}~\ref{fig:ocm-250-focal-plane} shows OCM and ISAM images of the full strongly scattering system with the focal plane placed deeper at $z_{\rm f}=250$ \textmu m, which does not improve the image quality.

\bibliography{suppl}